\documentclass[12pt]{article}

\usepackage{amsmath}
\usepackage{amssymb}
\usepackage{amsfonts}
\usepackage{graphicx}
\begin{document}

\begin{center}
{\large \bf{ On the neutron lifetime anomaly }}
\end{center}

\vspace*{1.0 cm}

\begin{center}
{
Paolo Cea~\protect\footnote{Electronic address:
{\tt paolo.cea@ba.infn.it}}  \\[0.5cm]
{\em INFN - Sezione di Bari, Via Amendola 173 - 70126 Bari,
Italy} }
\end{center}

\vspace*{1.0 cm}

\begin{abstract}
\noindent 
We suggest that the neutron lifetime anomaly can be  resolved by invoking the Casimir effect for  trapped
 ultracold neutrons. Our proposal can be checked experimentally by precise lifetime measurements of trapped ultracold
 neutrons with different trapping volumes.
\end{abstract}

\vspace*{0.6cm}

\vspace{0.2cm}
\newpage
\noindent
A free neutron decays into a proton, an electron and an antineutrino through the beta-decay process with a decay lifetime of about fifteen minutes.
This neutron decay is the prototype semileptonic weak decay and it constitutes the simplest example of nuclear beta decay.
The nucleon vector and axial vector weak coupling constants $G_V$ and $G_A$ determine the neutron lifetime together with the 
strengths of weak interaction processes involving free neutrons and protons that are of fundamental importance in Astrophysics, Cosmology,
Solar and Neutrino Physics~\cite{Dubbers:2011}. \\
The measure of the neutron lifetime has been implemented with two different experimental strategies, namely the beam method and the bottle
method~\cite{Wietfeldt:2011,Wietfeldt:2018,Serebrov:2019a,Serebrov:2019b}. In beam experiments the rate of neutron decay is determined by
counting the decay products and the number of neutrons in a well-defined volume of a neutron beam. In the bottle experiments neutrons
with very low energy are confined in a trap or bottle established by material walls, magnetic fields and gravity. The lifetime is inferred from the measurements of the number of trapped ultracold neutrons that remain after a given decay period.
Unfortunately, the two experimental methods for measuring the neutron lifetime disagree. In fact, in Ref.~\cite{Wietfeldt:2018} it is
reported a summary of neutron lifetime experimental measurements since 1985. The weighted average of the beam-method lifetime is:
\begin{equation}
\label{1}
\tau_N^{beam} \; = \; 888.1 \; \pm \; 2.0 \; \; s \; \; \; ,
\end{equation}
while  the bottle-method lifetime is:
\begin{equation}
\label{2}
\tau_N^{bottle} \; = \; 879.45 \; \pm \; 0.58 \; \; s \; \; \; .
\end{equation}
Assuming that the errors are Gaussian and combining errors in quadrature, we find:
\begin{equation}
\label{3}
 \Delta \tau_N \; = \; \tau_N^{beam} \; - \;    \tau_N^{bottle} \; = \; 8.65 \; \pm \; 2.08 \; \; s \; \; \; .
\end{equation}
We see that  the experimental measurements of the neutron lifetime are significantly different depending on the method since
the discrepancy is at level of almost four standard deviations. This discrepancy is mentioned in the scientific literature 
as {\it neutron lifetime anomaly}\footnote{For a critical discussion see Ref.~\cite{Serebrov:2021}.}.
 Actually, there is a lively discussion that the discrepancy comes
from unconsidered systematic errors or undetectable decay modes. \\
We believe that the neutron lifetime anomaly points to well-defined physical effects that have been overlooked until now. Indeed,
in this note we advance the proposal that the discrepancy can be fully explained by taking into account the 
Casimir effect~\cite{Plumien:1986,Milton:2001,Bordag:2009,Dalvit:2011,Simpson:2015} in the theoretical  estimate of the lifetime of
trapped ultracold neutrons (the bottle method). 
To this end, we consider the transition amplitude relevant for the neutron
 beta-decay\footnote{We shall follow the conventions adopted in Ref.~\cite{Wietfeldt:2011}.}:
\begin{equation}
\label{4}
{\mathcal{M} } \,  = \,  \left [ G_V  \bar{\Psi}_{P}  \gamma_{\mu}  \Psi_{N}  - G_A  \bar{\Psi}_{P} \gamma_5  \gamma_{\mu}  \Psi_{N} \right ]  
 \left [  \bar{\Psi}_{e}  \gamma^{\mu} (1 + \gamma_5) \Psi_{\nu_e}  \right ]  \; ,
\end{equation}
where we recall that $G_V$ and $G_A$ give the strengths of the vector and axial  vector couplings. The neutron decay probabilitry
per unit time is given by:
\begin{equation}
\label{5}
 d W \,  =  \, \frac{| {\mathcal{M} }|^2 }{(2 \pi)^5} \; \delta(E_e + E_{\nu_e} - \Delta) \,  \frac{1}{2 E_e} \frac{1}{2 E_{\nu_e}}  \,
 d^2 p_{\nu_e} d^3 p_e   \; ,
\end{equation}
were:
\begin{equation}
\label{6}
\Delta  \;  =  \;  m_N \; - \; m_P  \;  \; . 
\end{equation}
Integrating over the antineutrino momentum and electron solid angle one gets the electron energy spectrum:
\begin{equation}
\label{7}
 \frac{ d W }{d E_e} \;   =  \;  \frac{G_V^2 + 3 \, G_A^2}{2 \pi^3} \; E_e \, |\vec{p_e}|  \, ( \Delta - E_e)^2    \;   \; .
\end{equation}
Further integration over the electron energy gives the exponential decay constant:
\begin{equation}
\label{8}
 \frac{1}{\tau_N} \; = \; W \;   =  \;  \frac{G_V^2 + 3 \, G_A^2}{2 \pi^3} \; f_R( \frac{ \Delta}{m_e}) \;  m_e^5    
\end{equation}
where $f_R(x)$ is the  integrated Fermi phase-space factor. The subscript $R$ indicates that the value of the integral over
the energy spectrum includes the Coulomb, recoil order and radiative corrections. To illustrate our proposal we shall
neglect the higher-order corrections and, thereby, approximate the Fermi phase-space factor to the lowest order:
\begin{equation}
\label{9}
f_R(x)  \;   \simeq  \;   \frac{\sqrt{x^2 -1}}{60} (2 x^4 - 9 x^2 -8) +  \frac{x}{4} \log ( \sqrt{x^2 -1} + x ) \; \; .
\end{equation}
Note that, within our approximations,  Eq.~(\ref{8}) overestimates the lifetime of free neutrons by about 10 \%.
It is generally assumed that the above theoretical calculations should be  applied to both the beam and bottle-method
lifetime measurements. However, in the bottle method the ultracold neutrons are confined in a trap. It is known that quantum
fields in presence of external constraints, i.e. interaction with matter or external constraints, are led to modified zero-point fluctuations.
As a consequence, the vacuum energy density suffers a finite shift that can be considered as arising due to the change in the
zero-point energy of the quantum field (the Casimir effect). We may consider the trapped neutrons as quantum Dirac fields confined
in a bag with boundary conditions analogous to the MIT bag model~\cite{Johnson:1975,Hasenfratz:1978,DeTar:1983}.
Assuming that the Dirac quantum field is confined in a spherical cavity with radius $R$ and neglecting the effects due to the fermion 
mass one obtains~\cite{Johnson:1975}:
\begin{equation}
\label{10}
\Delta E_{vacuum} \; \simeq \; - \; \frac{7 \, \pi^3}{2160} \; \frac{1}{R}   \;  =  \; -  \;  \delta_C(R) \;  \; ,
\end{equation}
where $\Delta E_{vacuum}$ is the variation of the vacuum energy  due to the presence of the spherical bag. Note
that for $R \rightarrow \infty$ one recovers the case of quantum fields without external constraints. An immediate consequence of
Eq.~(\ref{10}) is that the energy of the excitations over the vacuum is  shifted by $ + \delta_C(R)$ with respect to the case
without external constraints. Therefore, in this case the neutron lifetime is given by:
\begin{equation}
\label{11}
 \frac{1}{\tau_N(R)} \;    =  \;  \frac{G_V^2 + 3 \, G_A^2}{2 \pi^3} \; f_R( \frac{ \Delta +\delta_C(R) }{m_e}) \;  m_e^5    \; \; .
\end{equation}
Our main result is that Eq.~(\ref{11}) should be applied to evaluate the lifetime for trapped neutrons (bottle method), while
Eq.~(\ref{8}) is relevant for the beam method. Introducing:  
\begin{equation}
\label{12}
 \Delta \tau_N(R) \;    =  \;  \tau_N(R  \rightarrow \infty) \; - \;  \tau_N(R)     \; \; ,
\end{equation}
we may attempt a comparison with observations.
\begin{figure}[t]
\centering
\includegraphics[width=0.80\textwidth,clip]{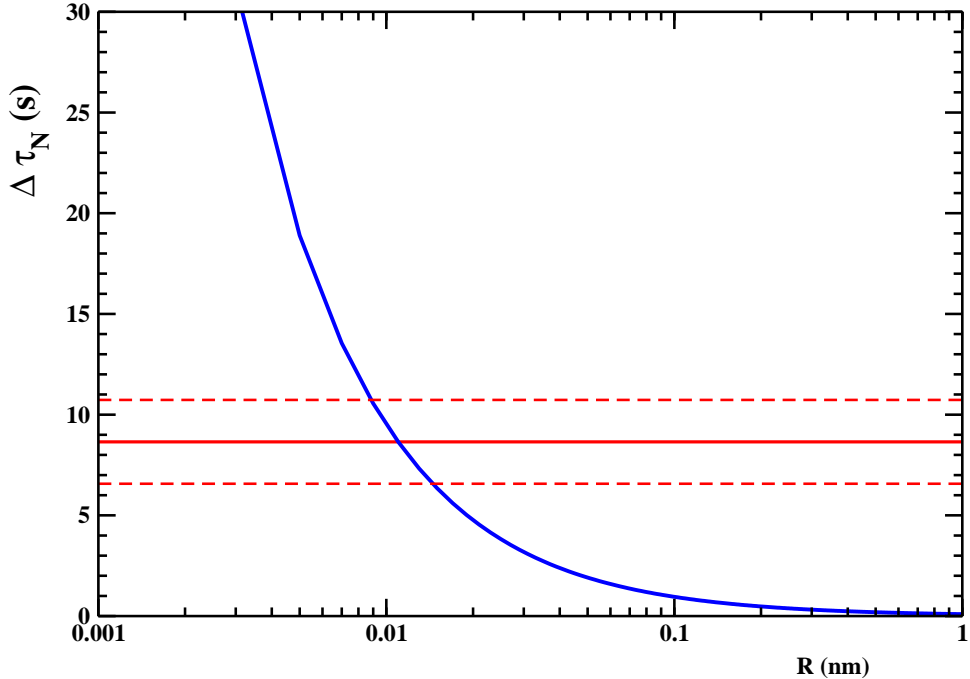}
\caption{\label{Fig1} Plot of $\Delta \tau_N$, Eq.~(\ref{12}), as a function of $R$ (blu line). The horizontal red line is the experimental value, 
Eq.~(\ref{3}), with one standard deviation uncertainty (dashed red lines).}
\end{figure}
 In fact, in Fig.~\ref{Fig1} we compare  $\Delta \tau_N$ given by Eq.~(\ref{3}) with our theoretical estimate Eq.~(\ref{12}). \\
 For the numerical calculations we used:
\begin{equation}
\label{13}
G_V \;    =  \;  G_F \, V_{ud}  \; \; , \; \;   G_A \; = \; \lambda \, G_V
\end{equation}
where  $G_F$ is the Fermi coupling constant~\cite{PDB:2020}:
\begin{equation}
\label{14}
 G_F  \;  \simeq  \;  1.16638 \, \times \, 10^{-5}  \; \text{GeV}^{-2}      \; \; ,
\end{equation}
$V_{ud}$ the CKM quark mixing matrix element:
\begin{equation}
\label{15}
 V_{ud}  \;  \simeq  \;  0.97370       \; \; ,
\end{equation}
and
\begin{equation}
\label{16}
\lambda  \;  \simeq  \;  - \, 1.2756       \; \; .
\end{equation}
Looking at Fig.~\ref{Fig1} we see that the neutron lifetime anomaly can be accounted for if we assume $R \simeq 10^{-9}$ cm.  \\
A few comments are in order. We have considered the effect due to the vacuum fluctuations of trapped ultracold neutrons. We approximated 
neutrons as free quantum Dirac fields and neglected the neutron mass. Within these approximations the main effect due to the trapping volume
is a finite shift of the excitation energies. As a result the trap cavity modifies the phase-space density of states in the neutron beta-decays
leading to an increase of the transition rate. Actually, we should also consider the eventual shift corrections caused by the fluctuations of
the cavity electromagnetic modes. Indeed, in the precise measurements of the electron magnetic moment the electrons are confined in an 
ideal Penning trap (see, e.g.,  Ref.~\cite{Gabrielse:2010} and references therein). To reach the desired precision in the measurement of
$g_e/2$ one needs to take care of the corrections that depend upon the interaction of the electrons with the cavity electromagnetic
radiation modes. It turned out~\cite{Boulware:1985} that there are no significant correction to the spin-precession frequency, but
the correction to the cyclotron frequency may be important. Therefore, to extract the electron magnetic moment it is necessary to
isolate the cavity shift that arises from the interaction of the cyclotron motion and the trap cavity~\cite{Gabrielse:2010}.
Now, since the neutrons are neutral spin-1/2 particles, the main interaction with the electromagnetic field is through the magnetic
moment which is proportional to the spin. Then, we see that we can safely neglect the influence of the cavity electromagnetic quantum
fluctuations. \\
The most important aspect of our proposal is that the lifetime of trapped ultracold neutrons does depend on the
size of the confining  trap.  Even though in the bottle method the experimental determination of the neutron trapping volume
is quite difficult, our estimate of the radius $R$ of the spherical ball is likely smaller than the geometrical linear size of the trapping
cell. It should be stressed, however, that due to our oversimplifications  this estimate of the radius of the spherical bag must be considered
with caution. Then, we must consider our parameter $R$ as an effective radius that, obviously, should depend on the trapping size.
We do not known if the effect we are discussing in the present paper  may account for the neutron lifetime anomaly. 
Nevertheless,  in principle, it is possible to check experimentally our proposal by performing precise
enough  measurements of $\tau_N^{bottle}$ using traps with varying volumes. Actually, in Ref.~\cite{Serebrov:2019b} it is reported that
there are different bottle experiments that are in agreement. However, this does not necessarily ruled out our proposal. In fact,
it could well be that experimental set-ups with different trap geometries could be characterised by comparable  $R$.
From Fig.~\ref{Fig1} we infer that, in order to detect a clear experimental signature of our Casimir effect,  one should increase
the effective trap size $R$ by almost a factor of two thereby increasing the trapping volume by about an order of magnitude.
From the experimental point of view this seems to be rather difficult. For instance,  in experiments adopting the  ultracold neutron magnetic traps
the manufacturing of a large magnetic trap is quite problematic due to severe limitations of the presently available technology. \\
Finally,  it is important to mention that another important consequence of our proposal resides on the fact that the correct neutron lifetime to be used 
in phenomenological analyses should be the beam-method lifetime, Eq.~({\ref{1}).

\end{document}